\documentstyle[12pt]{article}

\newcommand{\ftn}{\footnotesize}
\begin{document}
\begin{titlepage}
\title{Discretization on the Cosmic Scale Inspired from 
the 
Old Quantum Mechanics}

\author{A. G. Agnese\thanks{
Universit\`a di Genova,Dipartimento di Fisica, and 
INFN, Sezione di Genova,\newline
Via Dodecaneso 33, 16146 Genova (Italy). 
E-Mail: agnese@genova.infn.it .} \and
R. Festa\thanks{
Universit\`a di Genova,Dipartimento di Fisica, and 
INFM, Unit\`a di Genova,\newline
Via Dodecaneso 33, 16146 Genova (Italy). 
E-Mail: festa@genova.infn.it .}
}
\date{\small To appear in: {\sl Proceedings of the Beer Sheva Workshop
on Modern Modified Theories of Gravitation and Cosmology}~~(1997)}
\maketitle
 
\begin{abstract}
The Old Quantum Mechanics action discretization rules for periodic motions
on the atomic scale (Bohr-Sommerfeld) have been suitably modified in order to
take into account the gravitational field instead of the electrostatic 
one. The new rules are used to calculate a few mechanical
quantities pertinent to the periodic motions of celestial objects, 
and several values are obtained 
which result in reasonable agreement with the corresponding experimental data.

\noindent A gravitational dimensionless structure constant $\alpha_g$ is
determined, using the data relative to the solar system, which
allows to quantitatively account for phenomena on a much wider
scale. In particular, some information is acquired 
about the recently discovered
extrasolar planetary systems and about the general empirical law
which connects the spin of a celestial body with the square of its mass. 

\noindent Though no general theory is explicitely proposed in order to explain 
the obtained numerical coincidences, Nelson's stochastic mechanics
and/or chaotic dynamics seem to be the best candidates for this r\^ole.
\end{abstract}
 
\end{titlepage}

\noindent {\bf I. Introduction}

In a recent work \cite{Clues} we suggested that the periodic motions
of celestial bodies can be described, approximately enough, starting from 
suitably   tuned action discretization rules (scaling up of the
Bohr-Sommerfeld laws).

\noindent In fact, if the Planck's constant is expressed using 
the relationship 
\[
h = 2\pi\;\frac{e^2}{\alpha_e c}\; , 
\]
where $\alpha_e \simeq 1/137$ is the dimensionless fine structure 
constant, the 
Bohr-Sommerfeld rules can be written in the form
\[
\oint p_j\;dq_j = n_j \; 2 \pi \frac{e^2}{\alpha_e c}\;.
\]
One can imagine that,
if the electron charge was known with higher precision in Planck's time,
he would have introduced $\alpha_e$ instead of 
$h$ as a new constant.

The form of the previous equation explicitely shows its 
`atomic electrostatic' 
purpose. On the other hand, since both 
the (atomic) electrostatic and the gravitational force field follow the
$r^{-2}$ law and only differ in their `sources':
\[
\begin{array}{lcl}
e^2 &=&\mbox{\rm proton charge times electron charge in the H atom}\\
GMm &=&\mbox{\rm mass m in the gravitational field of a source of mass M ,}
\end{array}
\]
we will conjecture that the Old Quantum Mechanics rules hold if one performs
the substitution
\[
\frac{e^2}{\alpha_e}\longrightarrow\frac{G M m}{\alpha_g}\;,
\]
where $\alpha_g$ is a (new) dimensionless `gravitational' structure constant
to be determined.

Introducing the gravitational potential energy
\[
V = -\frac{GMm}{r}\;,
\]
it follows by simple calculations that the major 
semi-axes of the (elliptical)
orbits are given by
\[
a_n = n^2 a_1\;,
\]
where $n=1, 2,\ldots$ is the principal number and
\[
a_1 = \frac{G M}{\alpha_g^2 c^2}
\]
is the `Bohr radius' of the $M$ gravitational source.

Other immediate consequences concern 
the minimum and maximum orbital distances
$q_{nl}$ and $Q_{nl}$ and the corresponding maximum and minimum speeds
$V_{nl}$ and $v_{nl}$:
\[
\label{Sommernew}
\left\{
\begin{array}{ccccccc}
q_{n,l}& =& a_n (1-\varepsilon)&\hspace{0.5cm}
                               ;\hspace{0.5cm}&V_{n,l}&=&
(1+\varepsilon) \frac{\alpha_g c}{l}\\
Q_{n,l}& =& a_n (1+\varepsilon)&\hspace{0.5cm}
                               ;\hspace{0.5cm}&v_{n,l}&=&
(1-\varepsilon) \frac{\alpha_g c}{l}\\ 
\end{array}\right.\;,
\]
where 
\[
\label{eccentricity}
\varepsilon = \sqrt{1-\frac{l^2}{n^2}}
\] 
is the elliptical eccentricity, and,
as customary, the azimuthal quantum number $l$ can take the values
$1,\ldots,n$.
\noindent The angular momentum is given by
\[
j_{n,l} = l\;\frac{GMm}{\alpha_g c}\;.
\]
In the case of circular orbits ($l=n$) one has the constant speed
\[
v_n =\frac{\alpha_g c}{n}\;,
\]
the corresponding period
\[
P_n = 2 \pi\; \frac{G M}{\alpha_g^3  c^3}\; n^3\;,
\]
and the angular momentum
\[
j_n = \frac{GMm}{\alpha_g c} n\;.
\]

Thus, the universal constant
\[
v_* = \alpha_g c
\]
turns out to be the maximum speed for a gravitationally bounded
orbitating point mass. Correspondingly, 
\[
P_* = 2 \pi\; \frac{G M }{\alpha_g^3 c^3}\;, 
\]
which depends on the strength M of the gravitational source,
turns out to be its minimum period, and
\[
j_* = \frac{GMm}{\alpha_g c}
\]
its minimum angular momentum.

\vspace{0.5cm} 
\noindent{\bf II. A `domestic' test case: orbital motions around the Sun}

As a first test case we can apply the previous formulae to the motion
of the solar planets. In this case we use the following relationships
\[
\left\{
\begin{array}{lll}
\mbox{\rm Major semi-axes :} &a_n=n^2 a_1^\odot\;, & 
\mbox{\rm where  }a_1^\odot =\frac{G M^\odot}{\alpha_g^2 c^2}\\
  & & \\
\mbox{\rm Periods :}         &P_n=n^3 P_1^\odot\;, & 
\mbox{\rm where  }P_1^\odot =\frac{2\pi G M^\odot}{\alpha_g^3 c^3}\\
  & & \\
\mbox{\rm Mean speeds :}     &v_n =\frac{\alpha_g c}{n}&  
 
\end{array}
\right.
\]
where $M^\odot$ indicates the mass of the Sun.

In order to determine $\alpha_g$, we previously need to select
the quantum principal numbers for the
various planets. Note that, when a given number for a given planet
is chosen,
the numbers for all the other planets turn out to be determined
by their observed mean speeds.
Of course, the results are significant if the
concerned quantum principal numbers are small, 
and if no satisfactory agreement exists for
semi-integer n values.

One can easily verify that the only consistent choice lies in 
assigning to Mercury the principal number 3. Using the corresponding known
speed one obtains the initial guess for the gravitational structure constant
\[
\alpha_g = 3 \;\frac{v_{Mercury}}{c}\;,
\]
and the following quantum number selection:
Mercury$\rightarrow$3, Venus$\rightarrow$4, Earth$\rightarrow$5,
Mars$\rightarrow$6, Jupiter$\rightarrow$11, Saturn$\rightarrow$15,
Uranus$\rightarrow$21, Neptune$\rightarrow$26, Pluto$\rightarrow$30.

On this basis, one can use the relationship $v_n = \alpha_g c / n$ to refine
the estimate of
$\alpha_g$ through a general least square fit. 
 
In this way, the following results can be obtained:
\[
1/\alpha_g = 2086 \pm14\;,
\]
\[
v_* = 143.7 \pm 1.0\; km/s\;,
\]
\[
a_1^\odot = 0.04297\;a. u.\;,
\]
\[
P_1^\odot = 3.269\;d\;,
\]
and
\begin{center}
\begin{tabular}{||l||c|c|c|c||} \hline \hline
Planet       &$ n$ & $v_{obs}\; (km/s)$&$v_{calc}\; (km/s)$ \\  \hline
Mercury      & 3 & 47.87             &47.90          \\ 
Venus        & 4 & 35.02             &35.92          \\ 
Earth        & 5 & 29.78             &28.74          \\ 
Mars         & 6 & 24.13             &23.95          \\ 
Jupiter      & 11& 13.06             &13.06          \\
Saturn       & 15& 9.64              &9.58           \\
Uranus       & 21& 6.80              &6.84           \\
Neptune      & 26& 5.43              &5.53           \\ 
Pluto        & 30& 4.74              &4.79           \\ \hline\hline
\end{tabular}
\end{center}

In Fig. 1  the $v_n = \alpha_g c/n $ and $a_n = n^2 a_1$ 
relationships are
graphically compared  with the actual orbital mean speeds and the actual
major semi-axes of the 
planets of the Sun. One can see 
that the
solar planetary motions are reasonably accounted by
our model of `gravitational atom', even for small 
principal numbers.
 
As the azimuthal numbers are concerned,
considering one of the two planets with valuable eccentricity, 
i.e. Pluto ($n=30$),
and using the eccentricity formula with $l =29$, one obtains 
$\varepsilon_{calc} = 0.256$,
in fair agreement with the observed value $\varepsilon_{obs} \simeq 0.25$. 
Unfortunately, no allowable calculated value fits satisfactorily
the eccentricity of Mercury (whose motion shows on the other hand the
well known anomalies). 

Very recently \cite{1996TL} a new object has been discovered, named $1996TL_{66}$, 
which
can be assumed to adequately represent a number of bodies located
between the Kuiper belt (beyond the Neptune orbit) and the Oort cloud (which
reachs abouto 50,000 a.u. from the Sun).
The orbit of this object, the brightest trans-neptunian body but
Pluto and Charon, shows the very large eccentricity $\varepsilon = 0.58$
and a major semiaxis $a_n =\; 83.7 a.u.$. 

Using our mechanical model, one gets the $(n,a_n)$ pairs $(43, 79.4\;a.u)$,
$(44, 83.2\;a.u.)$ and $(45, 87.0\; a.u.)$: thus, the principal number
$n=44$ seems to be adequate to this `new planet'. `Scanning now the
allowable principal-azimuthal number pairs we obtain 
$(44, 35)\rightarrow\varepsilon=0.61$,
$(44, 36)\rightarrow\varepsilon=0.58$ and
$(44, 37)\rightarrow\varepsilon=0.54$, so that the pair $(n,l)=(44,36)$
can be assigned to $1996TL_{66}$.  Nevertheless, we explicitely agree
on the fact that results based on 
such large quantum numbers can not contribute to
confirm or disprove our model.
 
\vspace{0.5cm} 
\noindent{\bf III. Orbital motions around the planets}

\noindent On the basis of the assumed universality of the constant $\alpha_g$
one can use, in evaluating the `Bohr radius' of each planet, the formula
\[
r_1^{(planet)} = a_1^\odot \frac{M^{(planet)}}{M^\odot}\;,
\]
Through $r_1^{(planet)}$ one can easily calculate the allowed orbits relative to
each planet. In particular, one can determine the principal
number $n_f$ which corresponds to the first actual `free' orbit (i.e.,
to the allowed orbit immediately out of the planet body). 
In the following table
we show some interesting results (all the lengths are reported in  km, and
the general relative uncertainty of the calculated radii
is about $1\%$).
In the last two columns we report the greatest `not-free'
calculated orbital radius and the observed
equatorial radius of the concerned body.
Note that, in estimating $n_{free}$, the well known internal
stability Roche limit \cite{Smoluchowski}
has not been taken into account: some underestimation could result.
 
Considering the
principal number of the first 'free' orbit, one can note that
that for the internal planets (and for Pluto)
these orbits correspond to 
relatively high ($\ge 19$) principal numbers. 
The opposite, i.e. small ($\le 10$) `free'
quantum nubers, is verified
for the big external planets. 

\begin{center}
\begin{tabular}{||l||c|c|c|c|c|c||} \hline \hline
Body         & Bohr radius  &  $n_{free}$&$r_{n_{free}}$  &$r_{n_{free}-1}$ & Eq. radius  \\  \hline
Sun          &$6.50 \;10^6 $  &  1    &$(6.50\pm 0.06)\;10^6$ & &$6.96\;\;10^5$      \\\hline
Mercury      & $1.08$          & 48    &$2488$         &$2386 $&2439  \\ 
Venus        & $15.86$         & 20    &$6344$         &$5725 $& 6051 \\ 
Earth        & $19.45$         & 19    &$7021$         &$6302 $& 6378 \\ 
Mars         & $2.09 $         & 41    &$3513$         &$3344 $& 3397 \\ 
Jupiter      & $6190 $         & 4     &$99040$        &$55710$&71492 \\
Saturn       & $1853 $         & 6     &$66708$        &$46325$&60268 \\
Uranus       & $282  $         & 10    &$28200$        &$22842$&25559 \\
Neptune      & $333  $         & 9     &$26918$        &$21312$&24764 \\ 
Pluto        & $4400 \;10^{-4}$& 160   &$1126 $        &$1112 $&1123  \\ \hline\hline
\end{tabular}
\end{center}

If the orbits of the 
natural satellites of the planets are considered  
one really obtains  
ambiguous 
results.
For instance, only three of the four
greater Jupiter satellites confirm our rule.
This fact might denote
some failure of our model or, on the contrary, that the fourth satellite does not originate from the same Jupiter-material
as its companions.
A particular case is that of the Earth-Moon pair: we will consider this case
later.

As the Saturn rings are concerned, one obtains the following results, where
the unit lenght is $10^3 km$, and $IE$, $OE$ indicate respectively the internal and the
external edge of the considered ring:
\begin{center}
\begin{tabular}{||l||c|c|c|c|c|c|c|c|c||} \hline
Rings         &$D_{IE}$&$C_{IE}$&$B_{IE}$&Cassini&$A_{IE}$&F    &$G$& $E_{IE}$&$E_{OE}$ \\ \hline
$r_{obs}$     & 67.0   &74.5    &92.0    &119.8  &122.2   &140.4&170& 180     &480      \\
$n$           &6       &        & 7      &       &8       & 9   &   & 10      & 16      \\
$r_{calc}$    &66.7    &        &90.8    &       &118.6 &150.1  &   &$185.3 $ &$474.4 $ \\  
\hline
\end{tabular}
\end{center}
The uncertainties of the calculated values are all about $1\%$.
The calculated inner edge positions of most rings   
are in fair agreement with the corresponding experimental values.
Note that the boundary of the C ring is presumably determined by the
interactions between the charged dust particles and the 
planetary magnetic field, so that a purely gravitational theory does
not account for this fact.
Note also the remarkable coincidence 
between the calculated and the observed values of the 
boundaries of the broad
rarefied E ring.          
\newpage 
\vspace{0.5cm}
\noindent{\bf IV. Extrasolar planets}

During the last two years a few extrasolar planetary 
systems have been
discovered on the basis of the detected periodical anomalies in the motions
of their stars. 
Unfortunately, the mass of most the concerned stars are at present 
estimated with some inaccuracy. Almost all
the discovered planets, with masses ranging between one half and ten times
the Jupiter mass,  turn out to move somewhat near to the star, 
their guessed
major semi-axes ranging from about $0.05\; a.u.$ and $1\; a.u.$ ( only
three planets show a mean distance from the central star greater
than $1\; a.u.$ and less than $2\; a.u.$).

In order to test our scheme, we have considered only the 
19 extra-solar planets which have been confirmed (at present) \cite{Internet} and
two more planetary systems (Proxima Centauri and Barnard's) 
whose star masses are known with sufficient accuracy. 
The star masses 
of the other 19 confirmed planetary systems have been guessed from the star type
information using the values $M_{st}$ reported in literature (\cite{Landolt2}). 
Then, the major semi-axes 
$a_{st}$ of
the planetary orbits have been calculated from the corresponding
measured periods.
Unfortunately, the starting point (i.e., the values of the star masses) 
was somewhat inaccurate.

Thus, an alternative path has been devised: starting from the measured planetary
periods, and using the relationship
\[
M_n = M_\odot \frac{P}{P_\odot} \frac{1}{n^3}\;,
\]
we scanned the various hypothetical star masses $M_1$, $M_2$, \ldots,
$M_n$ which would correspond to the various planet principal numbers $1, 2, 
\ldots,n$.
Then, the mass value $M_j$ nearest to the guessed star-type
mass $M_{st}$ was singled out: thus, the
the appropriate principal number $j$ was determined , 
and this in turn allowed
an appropriate estimate $a_j^{(est)}$ of the major semi-axis. 
In the following table, in order to put in evidence
the ambiguity level in choosing the `most likely' star mass, we report,
together the chosen mass $M_j$, its 
neighbouring hypothetical
values $M_{j-1}$ and $M_{j+1}$.

\noindent We explicitely note 
that for most of the considered planetary systems 
the choice of the appropriate $M_j$ is unequivocal enough.

For the last two planetary systems, the planet period
was astrometrically determined. In both cases, the chosen $M_j$ value
turns out to fit exactly the measured one: this fact seems to support the
correctness of our estimation method.

\begin{center}
\begin{tabular}{||l||c|c|c|c|c|c|c|c|c|c||} \hline
\ftn Star       &\ftn  Type  &\ftn  $M_{st}$ &\ftn P$^{(obs)}$  &\ftn $a_{st}$ &\ftn M$_{j-1}$&\ftn M$_j$  &\ftn M$_{j+1}$&\ftn n  &\ftn a$_j^{(est)}$\\ \hline
\ftn 51 Peg     &\ftn G2IVA  &\ftn 1.05      &\ftn 4.229        &\ftn 0.050    &\ftn  -       &\ftn 1.30   &\ftn 0.16     &\ftn 1  &\ftn 0.056\\
\ftn ups Androm.&\ftn F7V    &\ftn 1.20      &\ftn 4.611        &\ftn 0.057    &\ftn  -       &\ftn 1.42   &\ftn 0.18     &\ftn 1  &\ftn 0.061\\
\ftn 55 Cancer  &\ftn G8V    &\ftn 0.90      &\ftn 14.648       &\ftn 0.110    &\ftn 4.50     &\ftn 0.56   &\ftn 0.17     &\ftn 2  &\ftn 0.097\\
\ftn rho CrB    &\ftn G0V-G2V&\ftn 1.05      &\ftn 39.645       &\ftn 0.230    &\ftn 12.18    &\ftn 1.52   &\ftn 0.45     &\ftn 2  &\ftn 0.261\\
\ftn 16 Cyg B   &\ftn G2.5V  &\ftn 1.05      &\ftn 804.000      &\ftn 1.720    &\ftn 1.98     &\ftn 1.14   &\ftn 0.72     &\ftn 6  &\ftn 1.766\\
\ftn 47 Uma     &\ftn G0V    &\ftn 1.05      &\ftn 1088.445     &\ftn 2.110    &\ftn 1.55     &\ftn 0.98   &\ftn 0.65     &\ftn 7  &\ftn 2.050\\
\ftn tau Bootis &\ftn F6IV   &\ftn 1.30      &\ftn 3.313        &\ftn 0.046    &\ftn  -       &\ftn 1.02   &\ftn 0.13     &\ftn 1  &\ftn 0.044\\
\ftn 70 Virgo   &\ftn G4V    &\ftn 0.90      &\ftn 116.600      &\ftn 0.430    &\ftn 4.48     &\ftn 1.33   &\ftn 0.56     &\ftn 3  &\ftn 0.512\\
\ftn HD 114762  &\ftn F9V    &\ftn 1.20      &\ftn 84.050       &\ftn 0.300    &\ftn 3.23     &\ftn 0.96   &\ftn 0.40     &\ftn 3  &\ftn 0.369\\
\ftn HD 110833  &\ftn K3V    &\ftn 0.73      &\ftn 270.040      &\ftn 0.800    &\ftn 1.30     &\ftn 0.66   &\ftn 0.38     &\ftn 5  &\ftn 0.712\\
\ftn BD-04 782  &\ftn K5V    &\ftn 0.67      &\ftn 240.920      &\ftn 0.700    &\ftn 1.16     &\ftn 0.59   &\ftn 0.34     &\ftn 5  &\ftn 0.635\\
\ftn HD 112758  &\ftn K0V    &\ftn 0.79      &\ftn 103.220      &\ftn 0.350    &\ftn 1.17     &\ftn 0.50   &\ftn 0.25     &\ftn 4  &\ftn 0.340\\
\ftn HD 98230   &\ftn F8.5V  &\ftn 1.30      &\ftn 3.980        &\ftn 0.060    &\ftn   -      &\ftn 1.22   &\ftn 0.15     &\ftn 1  &\ftn 0.052\\
\ftn HD 18445   &\ftn K2V    &\ftn 0.73      &\ftn 554.670      &\ftn 0.900    &\ftn 1.36     &\ftn 0.79   &\ftn 0.50     &\ftn 6  &\ftn 1.219\\
\ftn HD 29587   &\ftn G2V    &\ftn 0.98      &\ftn 1157.843     &\ftn 2.500    &\ftn 1.65     &\ftn 1.04   &\ftn 0.69     &\ftn 7  &\ftn 2.180\\
\ftn HD 140913  &\ftn G0V    &\ftn 1.05      &\ftn 147.940      &\ftn 0.540    &\ftn 1.68     &\ftn 0.71   &\ftn 0.36     &\ftn 4  &\ftn 0.488\\
\ftn HD 283750  &\ftn K2     &\ftn 0.73      &\ftn 1.790        &\ftn 0.040    &\ftn  -       &\ftn 0.55   &\ftn 0.07     &\ftn 1  &\ftn 0.024\\
\ftn HD 217580  &\ftn K4V    &\ftn 0.70      &\ftn 454.660      &\ftn 1.000    &\ftn 1.12     &\ftn 0.65   &\ftn 0.41     &\ftn 6  &\ftn 0.999\\
\ftn Alpha Tau  &\ftn K5III  &\ftn 1.20      &\ftn 654.000      &\ftn 1.350    &\ftn 1.61     &\ftn 0.93   &\ftn 0.59     &\ftn 6  &\ftn 1.437\\
\hline
\ftn Prox.Cent. &            &\ftn 0.10      &\ftn 42           &              &0.20          &\ftn 0.10   &\ftn 0.06     &\ftn 5  &\ftn 0.111\\
\ftn Barnard's  &            &\ftn 0.12      &\ftn 132.0        &              &0.19          &\ftn 0.12   &\ftn  -       &\ftn 7  &\ftn 0.249 \\
\hline
\end{tabular}
\end{center}

\vspace{0.5cm} 
\noindent{\bf V. Quantized galactic redshifts}

Starting from the seventies, various authors (\cite{Tifft1},
\cite{Tifft2}, \cite{Arp1},\cite{Arp2},
\cite{GuthrieNapier}) found 
that the frequency analysis of the galactic redshifts (both
for binary galaxies and in the general case)
indicates speed differences which are
multiples of $37.5\;km/s$ (Tifft suggests $36\;km/s$).
Using our model, one can try to explain both this discretization 
and the estimated speed periodical structure as effects of
the speed quantization rule
\[
v_n = \frac{\alpha_g c}{n}=\frac{143.7}{n} \;km/s\;.
\]
 
\newpage
\vspace{0.5cm}  
\noindent{\bf VI. Spin of the celestial bodies}

In order to validate or disprove the effectiveness of our scheme and
the universality of the calculated $\alpha_g$,
we have taken consideration the known empirical
the relationship between the angular
momentum $J$ of a celestial body and its mass M.  
Studying this question, both from a theoretical and a phenomenological point
of view, Brosche (see for instance \cite{Brosche0}, \cite{Brosche1}, \cite{Brosche2}) and
later Wesson (\cite{Wesson1}, \cite{Wesson2}) suggested that the law
\[
J = p M^2
\]
holds for purely gravitationally bounded systems. Wesson, fitting 
the empirical values
and taking into account some theoretical consistency constraints suggested
to use the value
\[
p \simeq 8\;\; 10^{-16}\;cm^2\;g^{-1}\;s^{-1}.
\]

Consistently with the correspondence
$e^2 \longleftrightarrow G M m$ on which our mechanical model
is based, and recalling that the electron spin can be
written in the form
\[
s = \frac{1}{2} \frac{e^2}{\alpha_e c}\;,
\] 
we can suppose that the rotational momentum of a purely gravitationally 
bound celestial body is given by
\[
J =\frac{1}{2}\frac{GM^2}{\alpha_g c}\;.
\]
As a consequence, we get 
$p =G/(2\alpha_g c)= 2.32 \;\;10^{-15}\;cm^2\;g^{-1}\;s^{-1}$. 

We note that, independently from this new guess, the previous formula 
can also be directly obtained from our
considerations on the discretization of the orbital motions if 
a flat continuous purely gravitating disk is considered in which
each infinitesimal mass ring rotates on the Bohr radius
of the internal part of the disk which it delimits. In this case, the total
angular momentum can be easily calculated by the formula
\[
J = \int dJ =\frac{G}{\alpha_g c} \int m\;dm \;,
\]
which just gives the previously guessed expression.  
In fact, this remark suggests that the concerned expression 
works well for flat or almost flat celestial bodies (prevailing 
gravitational
energy) whereas some correction factor can be needed in the opposite cases
(when both the gravitational and the electrical energy play their role).
  
Anyway, we have compared these theoretical spin values with
empirical values found in the literature or, as in the case of most planets, 
with the values
\[
J_{obs} = \frac{2}{5}M R^2 \frac{2 \pi}{P_{obs}}\;,
\]
where $P_{obs}$ indicates the observed rotation period
(note that this formula refers to spherical and homogeneous 
rigid bodies and in fact it gives upper limits for the actual values).

In Fig.2 
we report the logarithmic graph 
of our expression $J=p M^2$  compared with the experimental data:
since no fit was performed,
the agreement between our formula and the data  
(over more than twenty orders of mass magnitude) can be judged quite good. 

In fact, the comparison shows some small discrepancy for the
the solar planets, 
which on the other hand are examples of celestial bodies held together by 
other forces besides the gravitational one.

In the following table the calculated and the "observed" 
(upper limits) angular momenta
for  some planets are shown: the Earth apart,
only for Saturn, owing to its oblate
shape, the angular momentum $J=8.16 \; 10^{37}\; J\;s$
can be directly calculated without using the previously remarked
assumptions about the spherical planet form and its mass distribution.

\noindent A clear discrepancy regards the Earth, which shows $J_{calc} =
8.24                                                                                                                                                                                                                                                                                                                                                                                                                                                                                                                                  \;10^{33} J\,s$ to be compared with $J_{obs} = 5.88\;10^{33}\,J\,s$.
This fact can probably be accounted by the permanent loss of angular momentum 
caused by the tidal friction. An increasing rate 
of the rotation period of the Earth of about 
$16\; s / 10^6 years \simeq 3\;10^{-12}$ has been  
estimated \cite{Smoluchowski}. This would correspond
to a Moon braking action beginning `only' about $1600 $ thousand millions of years
ago, long after the Earth ocean was formed.
On the other hand, the discrepancy in question
between calculated and measured angular momentum of
the Earth would be even
greater if the whole Earth-Moon system
($J_{obs} = 3.47\;10^{34}\,J\,s$) was considered. Both this and
the previous remarks seem to indicate
that
the Moon and the Earth can not be considered as a single celestial body:
both could have originated 
from protoplanetary material 
on the same orbital ring ($n=5$, $a_5 \simeq 1.07\; a.u.$)
and subsequently
have gravitationally captured each the other 
(actually the Earth captured the Moon) about $2\;10^9 \;years$ ago. 
This hypothesis would also account for
the $9\%$ of discrepancy between the observed and calculated terrestrial
orbital radius round the Sun.  

\begin{center}
\begin{tabular}{||l||c|c||} \hline
Body    & $J_{calc}\;\;$ [{\em J s}]& $J_{obs}\;\;$ [{\em J s}] \\  \hline
Mercury & $2.53 \;\;\; 10^{31}$ & $< 1.02\;\;\;10^{30}$ \\
Venus   & $5.50 \;\;\; 10^{33}$ & $< 2.14\;\;\;10^{31}$ \\
Earth   & $8.24 \;\;\;10^{33}$ & $5.88 \;\;\;10^{33}$ \\ 
Mars    & $9.56 \;\;\;10^{31} $&  $ < 2.10 \;\;\;10^{32}$ \\ 
Jupiter & $8.37 \;\;\;10^{38} $&  $< 6.83 \;\;\;10^{38}$ \\
Saturn  & $7.50 \;\;\;10^{37} $&  $ 8.10 \;\;\;10^{37}$  \\
Uranus  & $1.76 \;\;\;10^{36} $&  $<2.50 \;\;\;10^{36}$\\
Neptune & $2.46 \;\;\;10^{36} $&  $<2.30 \;\;\;10^{36}$ \\ 
Pluto   & $5.23 \;\;\; 10^{28} $&  $<1.5 \;\;\; 10^{29}$ \\ \hline 
\end{tabular}
\end{center}

\vspace{0.5cm}
\noindent{\bf VII. Conclusions}

\begin{itemize}
\item Assuming that 
the previously stated analogies hold between celestial and
Old Quantum Mechanichs,
one can calculate values for a number of celestial quantities which
turn out to be in remarkable agreement with the corresponding
observed values.

\item The gravitational structure constant $\alpha_g \simeq 1/2086$
has been used in order to account some observed
values related to
\begin{itemize}
\item the planetary motions around the Sun
\item the structure of the Saturn rings
\item the extra-solar planetary systems
\item the quantization of the galactic red-shifts
\item the spin of a broad class of celestial bodies, from solar planets to
supercluster.
\end{itemize}

\item No definite theory has yet been devised on the why 
the concerned analogies hold.
Nevertheless, Nelson's stochastic mechanics (\cite{Nelson},
 \cite{Blanchard})
seems at present the
best candidate to offer valuable suggestions. 
In fact, in Nelson's scheme, a particle of mass which moves
in a potential field is always subjected to a 
Brownian motion with diffusion coefficient
$\hbar / 2m $: thus, the particle motion can be described using
a probability density which turns out to obey 
the Schr\"odinger equation (of which the
rules of the Old Quantum mechanics are consequences). If one consider
the gravitational potential of a source of mass M, 
the diffusion coefficient of the appropriate
Brownian motion would be  $G M / (2 \alpha_g c) $. This approach has been
already applied to the study of the solar system, with some interesting
results (\cite{Albeverio1}, \cite{Albeverio2}).

\noindent In fact, we suspect that the Nelson's stochastic approach 
is an
effective way to describe a purely classical deterministic chaotic
dynamics, where the chaotic behaviour arises from the strong nonlinearities
in a many body problem. The verified existence of stable attracting orbits in
several many body problems can be a clue to the truth of this hypotesis.

A different interesting approach, 
which gives results very similar to those previously mentioned, 
has been worked out by Nottale (see for instance
\cite{Nottale1},\cite{Nottale2}). It is based on a scale
relativistic scheme, where the space-time resolution is an essential
physical variable, and the physical laws are to be scale-covariants
under resolution transformations: the fractal structure of the space-time
plays a r\^ole  of capital importance in the theory.
\end{itemize}

\vspace{0.5cm}
{\small
\noindent {\underline{\em Acknowledgments}}

\noindent We are much obliged to H. Arp 
for his kind encouragement and his interest in our work,
and for some
important piece of information he supplied
us about the work of both L. Nottale 
and N. Oliveira. We also express our thanks to P. Wesson for
his interesting suggestions and his pleasant call. One of us
thanks E.I. Guendelman for the kind hospitality in Beer Sheva.
}
\bibliography{discr.bib}
\bibliographystyle{unsrt}
\end{document}